\documentclass[twoside,a4paper,11pt]{proceedings}
\usepackage{graphicx}
\usepackage{hyperref}
\usepackage{movie15}
\usepackage{natbib}
\newcommand{\objsev}{\rm Cyg~OB2~\#7~}
\newcommand{\objeleven}{\rm Cyg~OB2~\#11~}
\newcommand{\Msun}{\ensuremath{\rm M_\odot}}

\newcommand{\ab}{Astrophysical Bulletin}
\newcommand{\apjjj}{ApJ}
\newcommand{\apjl}{Astrophys. J. Letters}
\newcommand{\mnras}{MNRAS}
\newcommand{\aaa}{A\&A}

\begin{document}
\pagenumbering{arabic}
\pagestyle{myheadings}
\thispagestyle{empty}
\vspace*{-1cm}
{\flushleft\includegraphics[width=8cm,viewport=0 -30 200 -20]{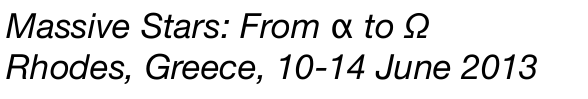}}
\vspace*{0.2cm}
\begin{flushleft}
{\bf {\LARGE
Modeling of atmospheres of the brightest stars belonging to the Cyg OB2 association
}\\
\vspace*{1cm}
Olga Maryeva$^1$, Valentina Klochkova$^1$ and Eugene Chentsov$^1$
%
}\\
\vspace*{0.5cm}
%
$^1$ Special Astrophysical Observatory RAS
%
\end{flushleft}
\markboth{
Modeling of atmospheres of the brightest stars belonging to the Cyg OB2 association
}{
Maryeva O.
}
\thispagestyle{empty}
\vspace*{0.4cm}
\begin{minipage}[l]{0.09\textwidth}
\ 
\end{minipage}
\begin{minipage}[r]{0.9\textwidth}
\vspace{1cm}
\section*{Abstract}{\small
                     We present the results of modeling of the spectrum of the brightest stars belonging 
                     to the Cyg OB2 association -- \objsev (${\rm O3If_*}$) and \#11 \  (${\rm O5Ifc}$). 
                     \objsev is one of the hottest stars in our Galaxy,  while \objeleven  belongs to 
                     the spectral class ${\rm Ofc}$, which was recently introduced and yet small in numbers \citep{WalbornOIfc}. 
                     We determine the physical properties and chemical composition of their atmosphere using 
                     CMFGEN code. The atmosphere of \objsev  reveals an excess of nitrogen $X(N_*)/X(N_\odot) = 3.2$ 
                     and the carbon and oxygen deficiency $X(C_*)/X(C_\odot) = 0.08$, $X(O_*)/X(O_\odot) = 0.09$, 
                     while the atmosphere of  \objeleven is  enriched with carbon $X(C_*)/X(C_\odot)\approx 1.$ 
                     and silicon $X(Si_*)/X(Si_\odot)\approx 2$. 

\vspace{10mm}
\normalsize}
\end{minipage}

\section{Introduction}


$\,\!$\indent       Cyg OB2 stellar association has been discovered in the middle of 20th century 
      and is still one of most often studied objects in the Galaxy. The association 
      is rich with young hot stars, and its total mass is estimated to be 30 000~$\Msun$. 
      The brightest stars belonging to the Cyg~OB2 association 
      have been investigated at the 6-m BTA telescope of the Special Astrophysical Observatory (SAO RAS) 
      since 2001 \citep{KlochkovaCygOB2,Chentsov}. It is worth noting that the О3-О5 supergiants 
      are the key objects for describing the evolution of this association. 
      We present the results of our study of two supergiants from Cyg OB2 --
      \objsev is   one of the hottest stars in our Galaxy, classified as ${\rm O3If_*}$ and 
      \objeleven \  classified as $ {\rm O5fc}$.

      Detailed studies of these stars are described in articles \citet{me,me2013,meobj11}.  
      Here we present only the main conclusions of these works.    


\section{Results}

     $\,\!$\indent  Using the non-LTE CMFGEN code \citep{Hillier5} we determined the physical 
     properties and  chemical composition of ${\rm Cyg~OB2~\#7~and \#11}$. 

\begin{figure*}[t]
      \center
      \includegraphics[width=0.8\columnwidth]{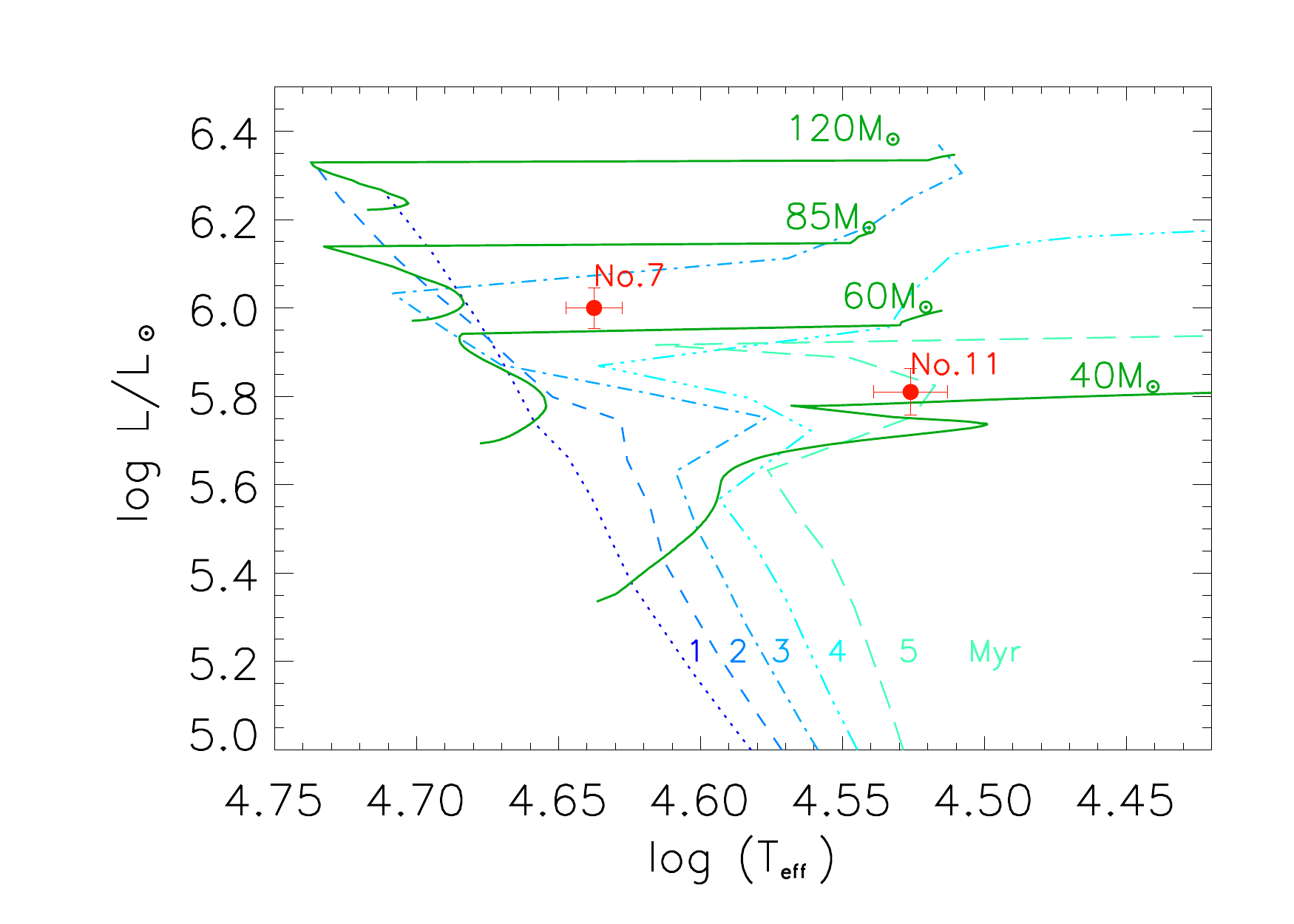}
      \caption{ The locations of \objsev \ and \#11 \  in the HR 
             diagram. The solid horizontal lines represent the mass tracks. 
             The vertical solid lines represent stellar isochrones. The evolution tracks 
             and the stellar isochrones are taken from the Geneva library.}
\label{fig:HRdiagram}
\end{figure*}   

     \objsev \ and \#11 are located in different parts of the association, 
     \#11 is in the outer part, \#7 lies near southern boundary of  the association  central part .
     Figure~\ref{fig:HRdiagram} shows the locations of these stars in the
     Hertzsprung-Russell diagram as well as evolutionary tracks and isochrones from
     the Geneva database \citep{Ekstrom}. The figure shows that the \objsev  lies 
     between the isochrones corresponding to three and  four Myr and mass$< 60~\Msun$, 
     while \#11 lies between four and five Myr and its mass is 40-60~\Msun. 
     \citet{Chentsov} spectroscopically confirmed age differences for stars which are located in different parts of the association. 
     The results of our numerical simulations support this conclusion and 
     are consistent with the hypothesis of  cascade  star formation in the association Cyg~OB2.

       Calculated abundances of the basic elements are given in Table~\ref{tab:frac}.
    The atmosphere of  \objeleven is enriched in carbon ($X_C/X_{C_\odot}=0.3-1.3$), silicium 
     ($X_{Si}/X_{Si_\odot} \simeq2.15-3$) and sulphur ($X_S/X_{S_\odot}\simeq 0.25$). Comparison of 
     \objeleven \ parameters with other supergiants shows that the high abundance of carbon itself does 
     not imply the $ {\rm OIfc} $ class. Moreover,  ${\rm Ofc}$ stars belonging to Cyg~OB2 association 
     are scattered in the  H-R diagram. Probably the appearance of strong   C{\scriptsize III}~$\lambda\lambda4647, 
     4650, 4652$~\AA \  emission lines in the spectra of O-stars is linked to the occurrence of  some physical 
     process rather than the chemical composition or evolutionary status. These are only assumptions 
     though, as we do not have enough statistics for reliable conclusions.

\begin{table*}[h]
\begin{center}
\caption{The abundances of chemical elements in the final model. X$_i$/X$_\odot$ -- is the ratio of the element abundance to the solar value. } 
\label{tab:frac}
\begin{tabular}{lccc||ccc}
           & \multicolumn{3}{c}{\objeleven}                                     &\multicolumn{3}{c}{\objsev}           \\
 Element   &   Fraction by the          &   Mass               & X$_i$/X$_\odot$  & Fraction by the   &   Mass            & X$_i$/X$_\odot$ \\
           &     number of atoms        &       fraction       &                  & number of atoms   &   fraction        &                 \\
\hline 
     H     &    $1.0$                   &    $0.82\div0.7   $  &  $1.2\div1.$        &     $1.0$         &    $0.5       $   &  $0.8 $         \\
     He    &    $0.05\div0.1$           &    $0.16\div0.28  $  &  $0.6\div1.$        &     $0.2$         &    $0.4       $   &  $1.6 $          \\
     C     &    $(1\div4)\times 10^{-4}$& $(1\div4)\times 10^{-3}  $  &  $0.3\div1.3$          &$3.8\times10^{-5}$ & $2.5\times10^{-4}$&  $0.08$         \\
     N     &$(1.7\div2.6)\times10^{-4}$ & $(2\div3)\times 10^{-3}  $  &  $1.8\div3.6 $         &$4.5\times10^{-4}$ & $3.5\times10^{-3}$&  $3.18$        \\
     O     &    $1.5\times 10^{-4}$     & $2\times 10^{-3}  $  &  $0.2$           &$9.75\times10^{-7}$& $8.6\times10^{-4}$&  $0.09$          \\
     Fe    &    $1.0\times 10^{-5}$     & $5\times 10^{-4}$    &  $0.37$          &                   &                   &                 \\
    Si     &    $6.5\times 10^{-5}$     & $1.5\times 10^{-3}$  &  $2.15$          &                   &                   &                 \\
     S     &    $3.5\times 10^{-6}$     & $9.3\times 10^{-5}$  &  $0.25$           &                   &                   &                 \\
\hline
\end{tabular}
\end{center}
\end{table*}

\small  
%
\section*{Acknowledgments}   
       Olga Maryeva thanks organizers of conference ``Massive stars from $\alpha$ to $\Omega$'' for a warm welcome and good organization!
       Also O.M would like to thank John D. Hillier for his great code {\sc cmfgen}, and 
       Nolan Walborn for valuable discussions. 
       The study was supported by the Russian Foundation 
       for Basic Research (projects no. 11-02-00319-a, 12-07-00739-a, 13-02-00029).   
      Olga Maryeva thanks the grant of Dynasty Foundation.

%

\bibliographystyle{aj}
\small
\bibliography{proceedings}

\begin{thebibliography}{}
\bibitem[Chentsov et al.(2013)]{Chentsov}
        E.L.~Chentsov, V.G.~Klochkova, V.E.~Panchuk, et al. 
Astron. Rep. \  {\bf 57}, Issue 7, 527 \ (2013).
\vspace*{-0.3cm}

\bibitem[Ekstr{\"o}m et al.(2012)]{Ekstrom}
{Ekstr{\"o}m}, S., {Georgy}, C., {Eggenberger}, P., et al. 
 \ 2012 \aaa,    \   537, A146
\vspace*{-0.3cm}
\bibitem[Hillier \& Miller(1998)]{Hillier5}
Hillier, D.J., Miller, D.L. 1998, ApJ, {\bf 496}, 407
 \vspace*{-0.3cm}

\bibitem[Kiminki et al.(2007)]{KiminkiAv}
{{Kiminki}, D.~C., {Kobulnicky}, H.~A., {Kinemuchi}, K.,  et al. 
}  
2007, \ \apjjj,   \   664, 1102 
\vspace*{-0.3cm}


\bibitem[Klochkova \& Chentsov(2004)]{KlochkovaCygOB2}
       Klochkova~V.~G. \& Chentsov~E.~L. \ Astron. Rep. \ {\bf 48}, 1005 \ (2004).

\vspace*{-0.3cm}
\bibitem[Kobulnicky et al.(2012)]{binaryO11}
{{Kobulnicky}, H.~A., {Smullen}, R.~A., {Kiminki}, D.~C., et al. 
}    
2012, \ \apjjj,  \ 756, 50

\vspace*{-0.3cm}
\bibitem[Maryeva et al.(2012)]{me}
{{Maryeva}, O.~V. \& {Zhuchkov}, R.~Y.} \ 2012, \ Astrophysics, 55, 371

 \vspace*{-0.3cm}
\bibitem[Maryeva {et al.}(2013a),Maryeva, Klochkova \& Chentsov]{me2013}
{{Maryeva}, O.~V.,   {Klochkova}, V.~G. \& {Chentsov}, E.~L.}
  \ 2013a, \ \ab,  \ 68,  87
\vspace*{-0.3cm}
\bibitem[Maryeva et al.(2013b)]{meobj11}
{{Maryeva}, O., {Zhuchkov}, R. \& Malogolovets, E.} \ 2013b, \ submitted to PASA

 \vspace*{-0.3cm}
\bibitem[Mel'nik \& Dambis(2009)]{Dambis}
{{Mel'Nik}, A.~M. \& {Dambis}, A.~K.} 2009, \ \mnras,  \    400, 518
\vspace*{-0.3cm}
\bibitem[Walborn et al.(2010)]{WalbornOIfc}
Walborn, N.~R., Sota, A., {Ma{\'{\i}}z Apell{\'a}niz}, et al. 
2010, \ \apjl,  \ 711, L143


\end{thebibliography}

\end{document}